\documentclass[12pt,preprint]{aastex} 
\usepackage[flushleft]{threeparttable}
\usepackage{graphicx}
\usepackage{epstopdf}

\def\gtorder{\mathrel{\raise.3ex\hbox{$>$}\mkern-14mu
             \lower0.6ex\hbox{$\sim$}}} 
\def\ltsima{$\; \buildrel < \over \sim \;$}
\def\simlt{\lower.5ex\hbox{\ltsima}}
\def\gtsima{$\; \buildrel > \over \sim \;$}
\def\simgt{\lower.5ex\hbox{\gtsima}} 

\begin{document} 


\title{The detection rate of early UV emission from supernovae:\\ A dedicated GALEX/PTF survey and calibrated theoretical estimates}


\author{Noam Ganot}
\affil{Department of Particle Physics and Astrophysics, Faculty of Physics, The Weizmann
Institute of Science, Rehovot 76100, Israel}

\author{Avishay Gal-Yam}
\affil{Department of Particle Physics and Astrophysics, Faculty of Physics, The Weizmann
Institute of Science, Rehovot 76100, Israel}
\email{avishay.gal-yam@weizmann.ac.il}

\author{Eran. O. Ofek}
\affil{Department of Particle Physics and Astrophysics, Faculty of Physics, The Weizmann
Institute of Science, Rehovot 76100, Israel}

\author{Ilan Sagiv}
\affil{Department of Particle Physics and Astrophysics, Faculty of Physics, The Weizmann
Institute of Science, Rehovot 76100, Israel}

\author{Eli Waxman}
\affil{Department of Particle Physics and Astrophysics, Faculty of Physics, The Weizmann
Institute of Science, Rehovot 76100, Israel}

\author{Ofer Lapid}
\affil{Department of Particle Physics and Astrophysics, Faculty of Physics, The Weizmann
Institute of Science, Rehovot 76100, Israel}

\author{Shrinivas R. Kulkarni}
\affil{Cahill Center for Astrophysics, California Institute of Technology, Pasadena, CA 91125, USA}

\author{Sagi Ben-Ami}
\affil{Smithsonian Astrophysical Observatory, Harvard-Smithsonian Ctr. for Astrophysics, 60 Garden St., Cambridge, MA 02138, USA}

\author{Mansi M. Kasliwal}
\affil{Cahill Center for Astrophysics, California Institute of Technology, Pasadena, CA 91125, USA}

\affil{(The ULTRASAT Science Team)}
\and

\author{Doron Chelouche}
\affil{Physics Department, Faculty of Natural Sciences, University of Haifa, 31905 Haifa. Israel}

\author{Stephen Rafter}
\affil{Physics Department, Faculty of Natural Sciences, University of Haifa, 31905 Haifa. Israel}

\author{Ehud Behar}
\affil{Physics Department, Technion Israel Institute of Technology, 32000 Haifa, Israel}

\author{Ari Laor}
\affil{Physics Department, Technion Israel Institute of Technology, 32000 Haifa, Israel}

\author{Dovi Poznanski}
\affil{School of Physics and Astronomy, Tel Aviv University, 69978 Tel Aviv, Israel}

\author{Udi Nakar}
\affil{School of Physics and Astronomy, Tel Aviv University, 69978 Tel Aviv, Israel}

\author{Dan Maoz}
\affil{School of Physics and Astronomy, Tel Aviv University, 69978 Tel Aviv, Israel}

\author{Benny Trakhtenbrot}
\affil{Zwicky Fellow; Institute for Astronomy, ETH Zurich, Wolfgang-Pauli-Strasse 27 Zurich 8093, Switzerland}

\affil{(The WTTH consortium)}

\and

\author{James D. Neill}
\affil{California Institute of Technology, 1200 East California Boulevard, MC 278-17, Pasadena, CA 91125, USA}

\author{Thomas A. Barlow}
\affil{California Institute of Technology, 1200 East California Boulevard, MC 278-17, Pasadena, CA 91125, USA}

\author{Christofer D. Martin}
\affil{California Institute of Technology, 1200 East California Boulevard, MC 278-17, Pasadena, CA 91125, USA}

\author{Suvi Gezari}
\affil{Department of Astronomy, University of Maryland, College Park, MD 20742-2421, USA}

\affil{(the GALEX Science Team)}

\and

\author{Iair Arcavi}
\affil{Las Cumbres Observatory Global Telescope, 6740 Cortona Drive, Suite 102, Goleta, CA 93111, USA and Kavli Institute for Theoretical Physics, University of California, Santa Barbara, CA 93106, USA}

\author{Joshua S. Bloom}
\affil{Department of Astronomy, University of California, Berkeley, CA 94720, USA}

\author{Peter E. Nugent}
\affil{Lawrence Berkeley National Laboratory, 1 Cyclotron Road, Berkeley, CA 94720, USA}

\author{Mark Sullivan}
\affil{School of Physics and Astronomy, University of Southampton, Southampton SO17 1BJ, UK}

\affil{(The Palomar Transient Factory)}



\begin{abstract} 
The radius and surface composition of an exploding massive star,
as well as the explosion energy per unit mass, can be measured 
using early UV observations of core collapse supernovae (SNe).
We present the first results from a simultaneous GALEX/PTF search for 
early UV emission from SNe. Six Type II SNe and one Type II superluminous
SN (SLSN-II) are 
clearly detected in the GALEX NUV data. We compare our detection rate with theoretical estimates
based on early, shock-cooling UV light curves calculated from models that fit existing
{\it Swift} and GALEX observations well, combined with volumetric SN rates.
We find that our observations are in good agreement with calculated rates 
assuming that red supergiants (RSGs) explode with fiducial 
radii of 500\,R$_{\odot}$, explosion energies of $10^{51}$\,erg, and
ejecta masses of 10\,M$_{\odot}$. Exploding
blue supergiants and Wolf-Rayet stars are poorly constrained. We describe
how such observations can be used to derive the progenitor radius, surface composition and 
explosion energy per unit mass of such SN events, and we demonstrate why
UV observations are critical for such measurements.
We use the fiducial RSG parameters to estimate the detection rate of SNe during the 
shock-cooling phase ($<1$\,d after explosion) for several ground-based surveys (PTF, ZTF, and LSST).
We show that the proposed wide-field UV explorer ULTRASAT mission, is expected to find
$>100$ SNe per year ($\sim0.5$ SN per deg$^{2}$),  independent of 
host galaxy extinction, down to an NUV detection
limit of $21.5$\,mag AB. Our pilot GALEX/PTF project thus convincingly 
demonstrates that a dedicated, systematic SN survey at the NUV 
band is a compelling method to study how massive stars end their life.
 
\end{abstract} 




\section{Introduction} 

Massive stars explosively end their life in a Core Collapse Supernova (CC SN). 
Few solid facts are known about SN progenitors. Hydrogen-rich Type II SNe (and in particular, Type II-P) are firmly
associated with red supergiant (RSG) progenitors, while rare underluminous SNe II (e.g., SN 1987A) may arise from blue 
superiants (BSG). Other classes of core-collapse SNe that are depleted in hydrogen (e.g., Types Ib, Ic)
probably arise from stripped stars, such as Wolf-Rayet (W-R) stars, but the exact mapping is unknown; see 
Filippenko (1997) and Smartt (2009) for reviews of SN types and progenitors, respectively. 
The final stages of massive star evolution and the physics of the explosion are
also poorly understood, see, e.g., Langer (2012) and references therein.

Although there are numerous
SN detections every year (Gal-Yam et al. 2013), most events are discovered a long time (days) after the explosion of the star.
This delay is unfortunate since
radiation emitted during the first few days after SN explosion is governed by relatively simple physics: recombination 
and line opacity are negligible and in most cases so is radioactivity. This early emission encodes crucial information about the outer envelope of the exploding star (approximately its outer $0.1$\,M$_{\odot}$) that can be extracted from robust and simple
models. Exploring this outer shell mass is very interesting as it is it that determines the stellar radius and outer density profile
of the star, and its properties can be used to study currently poorly-known stellar physics such as the mixing length and convection parameters. Observations
starting only after this early period thus result in loss of this information about the supernova progenitor star and the explosion
mechanism itself. Only a handful of events were detected during this early phase (e.g. Arnett et al. 1989;
Schmidt et al. 1993; Campana et al. 2006; 
Soderberg et al. 2008; Gezari et al. 2008,2010; Schawinski et al. 2008; Arcavi et al. 2011; Cao et al. 2013; Gal-Yam et al. 2011; 2014), and even in these cases the time resolution of the measurements is generally too poor to form a well-sampled light curve.

\begin{figure}
\includegraphics[width=15.5cm]{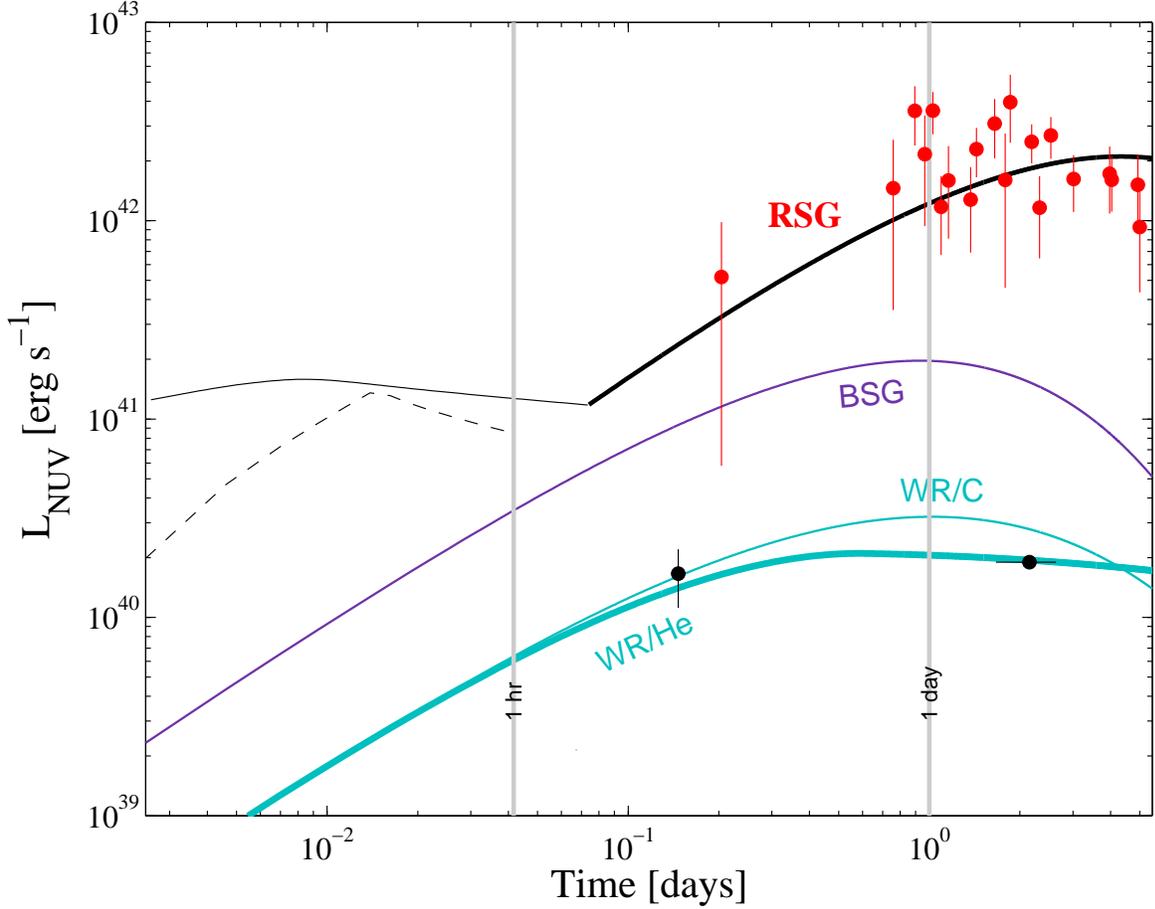}
\caption{SN early UV emission models compared to data. At early times ($<3$\,hr post explosion) no optical/UV
data exist. The models of
Sapir et al. (2013; solid) and Nakar \& Sari (2010; rescaled, see text; dashed) approximately agree in peak value, but differ in details. Forthcoming UV surveys (e.g., ULTRASAT; Sagiv et al. 2014) will observe such early emission and further constrain models. At
later times Rabinak \& Waxman 2011 (RW11; solid) models for red supergiant stars (RSG; thick black) and compact W-R stars 
(He, thick cyan and C/O, thin cyan) compare well with UV observations from {\it Swift}/UVOT (SN 2008D, Type Ib, Soderberg et al. 2008; black solid circles) and GALEX/NUV (SNLS-04D2dc, Type II,
Schawinski et al. 2008, Gezari et al. 2008, red solid circles). Blue supergiant
(BSG) models (thin blue) are currently untested. Stellar classes (RSG/BSG/WR) differ greatly in their UV peaks making 
early UV observations a strong discriminator among progenitor classes. 
Plotted models assume reasonable parameters: RSG with R$_{*}=500$\,R$_{\odot}$, explosion energy E$=2\times10^{51}$\,erg and ejected mass M$=10$\,M$_{\odot}$, BSG  with R$_{*}=50$\,R$_{\odot}$,  E$=10^{51}$\,erg and identical mass, and a W-R star with either He or C/O dominated composition, R$_{*}=1.15$\,R$_{\odot}$, E$=0.8\times10^{51}$\,erg and ejected
mass M$=7.5$\,M$_{\odot}$. RW11 models are unextinguished, data points have been extinction corrected (by A$_{NUV}=1.45$\,mag and
A$_{NUV}=2.2$\,mag for SNLS-04D2dc and SN 2008D, respectively) using the extinction values provided by Schawinsky et al. (2008) and
RW11 (for SN 2008D).}
\label{SBSCfig}
\end{figure}

An early detection of the SN and a measurement of its light curve are useful to understand the physics of
the explosion itself, and its progenitor properties.
The first light escaping from an exploding star emerges as a shock breakout flare, with 
a hot spectrum peaking in the ultra-violet (UV) or X-ray bands. Models for this shock breakout emission
have a long history (e.g., Colgate 1974; Grassberg et al. 1971; Falk 1978; Klein \& Chevalier 1978;
Ensman \& Burrows 1992, Matzner \& McKee 1999). 
In recent years several theoretical models were developed
in order to describe emerging observations of the explosion shock breakout (e.g., Nakar \& Sari 2010; 
Sapir et al. 2011; Katz et al. 2012; Sapir et al. 2013). Fig.~\ref{SBSCfig}
shows that exact analytic and numerical solutions by Sapir et al. (2013) are in general agreement with analytic 
models by Nakar \& Sari (2010) after appropriate rescaling\footnote{To correct the NS10 formulae we divided the luminosity
in their equations 29, 32 and 39 by a factor of 2.5. (E. Nakar; Personal Communication).} of the latter. 
However, only a single such flare has been serendipitously observed (Soderberg et al. 2008) and the relevant theory is 
virtually untested.  
If detected, shock breakout flares provide a direct measure of the 
pre-explosion stellar radius R$_{*}$: the flare duration scales as R$_{*}$/c, and the integrated luminosity as R$_{*}^{2}$
(Klein \& Chevalier 1978; Katz et al. 2012). 

Following an initial shock breakout flare (i.e., at times $\gtorder3$\,h post-explosion), 
the expanding stellar envelope emits a fraction of the leftover 
stored explosion energy during the shock cooling phase, initially peaking in the UV. This phase is better understood theoretically 
(e.g., Grassberg et al. 1971; Chevalier 1976, 1992; Chevalier \& Frensson 1998) and 
has been observed in a few cases (by GALEX, Schawinski et al. 2008, Gezari et al. 2008; and by {\it Swift},
Soderberg et al. 2008; data shown as red and black circles in Fig.~\ref{SBSCfig}). The shock cooling phase is longer and 
more luminous in larger stars. These works and, in particular,  
more recent models (e.g. Nakar \& Sari 2010, hereafter NS10; Rabinak \& Waxman 2011, hereafter RW11)
demonstrate that the shock breakout and subsequent cooling phases during the first days after explosion 
encode information about the SN progenitor radius and surface composition, the explosion energy per unit mass,
and the line of sight extinction (see below for details). 
This is strong motivation to design surveys targeting early UV emission from SNe. In this paper
we use the results of a pilot PTF/GALEX survey to robustly  
estimate the number of early SN detections expected from such surveys.

We review the derivation of physical progenitor and SN parameters from early observations of SNe in $\S~2$
and describe a sample of SNe with early UV emission detected by a GALEX/PTF wide-field experiment in $\S~3$.
We summarize our implementation of theoretical models in $\S~4$ and show that these fit the handful of available
data. We then combine these models with volumetric SN rates to estimate the expected number of detections
from the GALEX experiment we conducted in $\S~5$, show our fiducial models fit the observations well, 
and provide validated predictions for the proposed ULTRASAT space mission (Sagiv et al. 2014). We conclude
in $\S~6$.
 
\section{Motivation: deriving SN progenitor properties from early UV emission}

The early shock-cooling emission from SNe is governed by simple and well understood physics and can thus be used
to derive robust constraints on the physical parameters of the exploding star and of the explosion. Roughly, the rise-time
to peak determines the progenitor radius $R_*$, the peak flux determines the explosion energy per unit ejecta mass $E/M$, 
and the post-peak light curve constrains the surface composition $Z$ (RW11). This simple physics description holds as long as the temperature in the emitting region is $\gtrsim1$\,eV (see RW11 for details\footnote{Several complications, that prevent the construction of a simple and robust model, arise when the temperature of the emitting region drops below $\sim1$~eV (RW11): complicated opacity variations, significant contribution to the luminosity from recombination, and penetration of the photosphere
into deep envelope layers, which did not initially (i.e. before the explosion) lie at a very small distance, $dr_0<<R_*$, from
the surface of the star. As long as the emission is dominated by shells with $dr_0/R_*<<1$, the luminosity and the color
temperature are nearly independent of the pre-explosion density distribution. As the photosphere penetrates deeper,
the emission becomes dependent on the details of the density distribution (see the “+” signs in Figs. 2-4 of RW11, indicating
the limit of model validity).}), for which the emission peak is at $\lambda<0.3\mu$. In all optical bands (including the
$u$ band) the emission peak occurs only after the temperature falls below this threshold value (see Rubin et al. 2015 for 
detailed discussion). For this reason, the observational photometric parameters (rise-time
to peak, peak flux) cannot be related to physical parameters via a simple and robust model, making optical-light
observations not useful for this analysis.

For commonly assumed progenitor parameters, shock breakout is expected to be accompanied by soft ($0.3-10$\,keV) X-ray emission with luminosity of $10^{45}{\rm erg/s}$ ($10^{44}{\rm erg/s}$) for BSG (RSG/He-WR) progenitors (Sapir, Katz \& Waxman 2013). However, the ability to use X-ray observations to constrain progenitor and explosion parameters is limited by several factors.
 
(i)           First, the theory of X-ray emission from massive star explosions is not sufficiently well understood to ensure that stellar/explosion parameters can be reliably constrained based on X-ray observations. This is reflected, for example, by the fact that none of the few X-ray detections can be explained as shock breakout from a stellar edge (e.g., Sapir et al. 2013); these rather require more complex structures (such as winds or extended envelopes, e.g., Campana et al. 2006; Moriya et al. 2015).

(ii)          Second, the detection rate of X-ray breakouts is expected to be very low, even for future instruments with order of magnitude better sensitivities than past or current instruments. The non-detection of the predicted $10^{45}{\rm erg\,s^{-1}}$  soft X-ray breakout signal of BSG explosions (which are expected to dominate the detection event rate) in archival searches of ROSAT (Vikhlinin 1998) and XMM (Law et al. 2004) data imply an upper limit of $\sim10^{-7}\,{\rm\,Mpc^{-3}\,y^{-1}}$ on the rate of such events (Sapir et al. 2013; Sapir \& Halbertal 2014) which is about two orders of magnitude lower than the expected BSG explosion rate. This discrepancy may be related to the above mentioned tension between model predictions and observations, or to high obscuration of the explosions. In any case, it implies that a soft X-ray detector with a 1~sr FOV and sensitivity of $6\times10^{-11}{\rm erg\,s^{-1}}$ (over $\sim10$~s) will detect $<3$ events per year (consistent with the null detection of such events so far by MAXI; Camp et al. 2013). In a similar manner we can estimate the detection rate of early X-ray emission from SNe 
from the discovery of the early X-ray signal from
SN 2008D by Soderberg et al. (2008). SWIFT-XRT could detect SN 2008D-like events out to $200$\,Mpc (Soderberg 2008). 
 Even future wide-field Lobster telescopes will have a sensitivity which is $>100$ times less than XRT (Camp et al. 2013), and 
thus would detect such events only to $20$\,Mpc. This implies that even if the X-ray
breakout rate is as high as the entire core-collapse SN rate, $\sim10^{-4}$\,Mpc$^{-3}$\,y$^{-1}$, such a future mission
would detect 3 events per year (for an all sky detector).

(iii)         Finally, we note that massive star explosions associated with strong high-energy short transients, gamma-ray bursts (GRB) and X-ray flashes (XRF), like SN 2006aj (Campana et al. 2006), are both not understood theoretically and are very rare in the volumetric sense, as they account for $<< 1\%$ of core-collapse SNe (e.g., Podsiadlowski et al. 2004). The detection of such events cannot therefore be used to study the general properties of SN progenitor/explosion parameters.

\begin{figure}
\includegraphics[width=15cm]{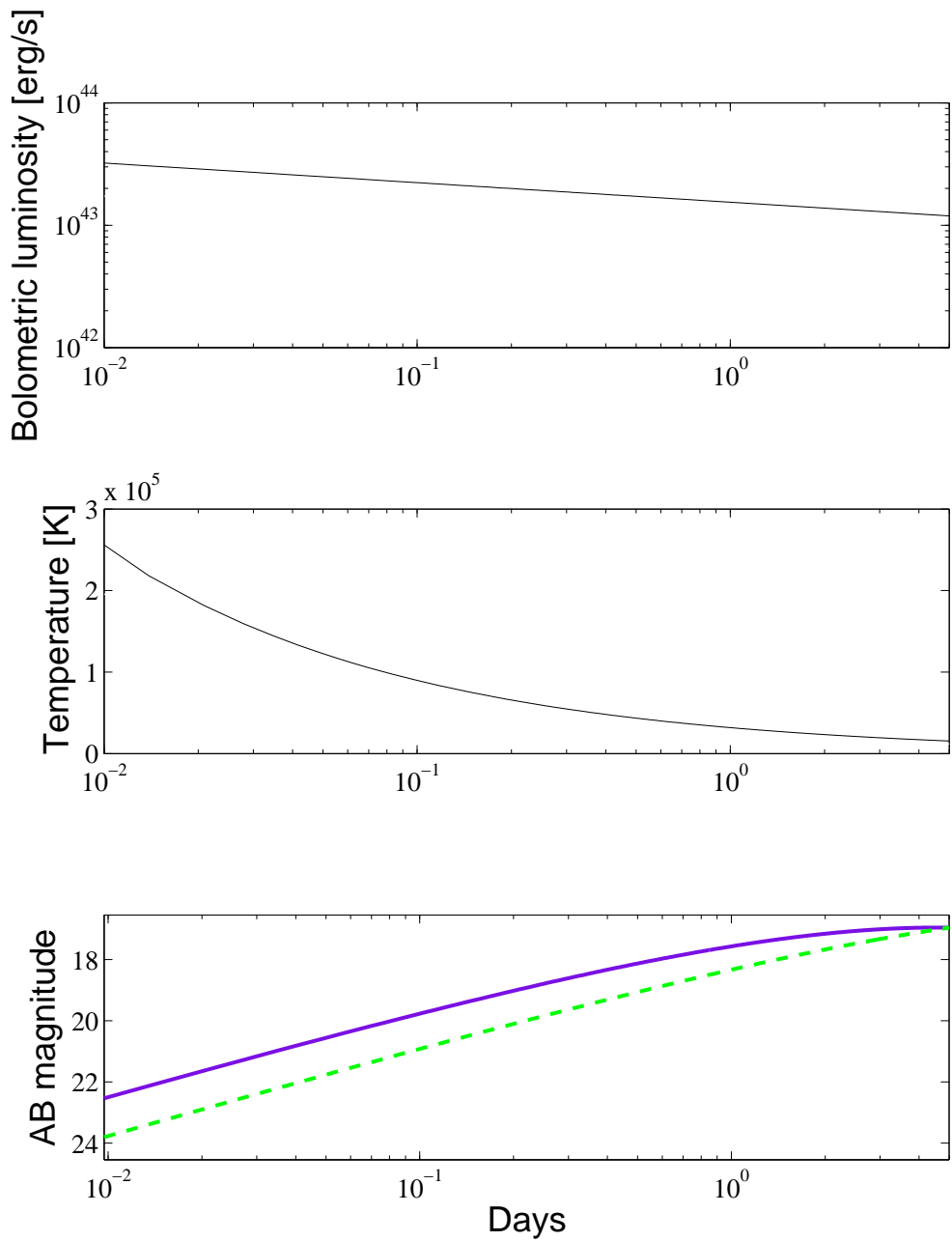}
\caption{The bolometric luminosity (top), temperature (middle) and NUV (purple) and $g$-band (green)
luminosities (bottom) predicted by the models of Rabinak \& Waxman (2011) for a fiducial RSG SN progenitor
with a radius of $500$\,R$_{\odot}$, explosion energy of $2\times10^{51}$\,erg, and ejected mass of 
$10$\,M$_{\odot}$. The rapid decline of the temperature leads to an NUV peak around 2 days
after explosion, when the black-body peak temperature crosses this band, 
while the optical $g$-band continues to rise beyond day 5. See animated version of this figure at
\protect\url{http://www.weizmann.ac.il/astrophysics/ultrasat/animations/ganot14a.gif}.
} 
\label{SCbasics}
\end{figure}

\begin{figure}
\includegraphics[width=17cm]{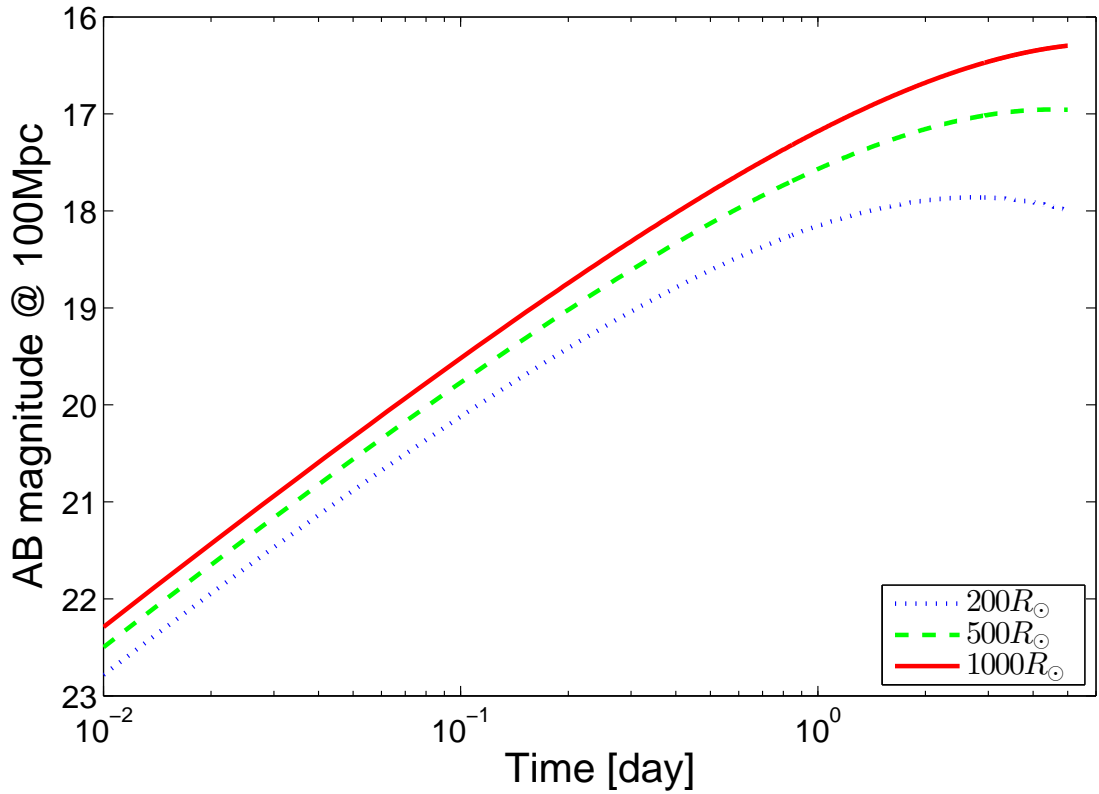}
\caption{RW11 Model SN NUV light curves for RSG explosions with identical parameters except for the radius R
(explosion energy E$=2\times10^{51}$\,erg, ejecta mass M$=10$\,M$_{\odot}$).
As can be seen, progenitor radii within the typical range for RSG stars ($200-1000$\,R$_{\odot}$) can be readily
distinguished by the light curve shape (time to peak). Note that this diagnostic is independent of the absolute scale
and so insensitive to extinction.}  
\label{getR}
\end{figure}

\begin{figure}
\includegraphics[width=17cm]{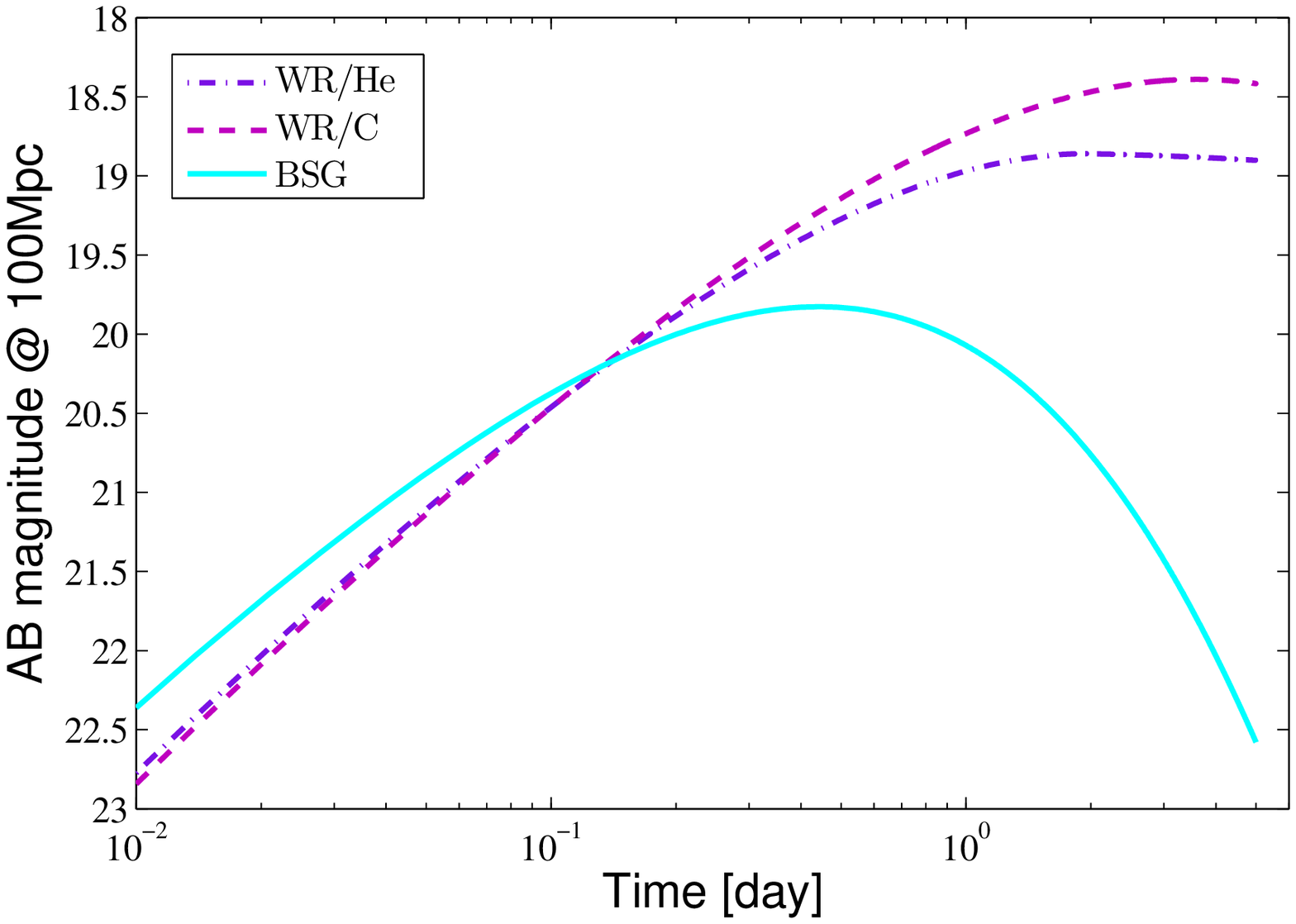}
\caption{Model RW11 SN NUV light curves for explosions of BSG (hydrogen envelope; cyan) and W-R (He and C/O envelopes; light and dark purple,
 respectively) progenitors with
identical radii ($10$\,R$_{\odot}$), explosion energy ($2\times 10^{51}$\,erg) and mass ($5$\,M$_{\odot}$). Well-sampled early
UV data ($<1$\,day) can readily diagnose both the radius and composition of compact stars.}
\label{getZ}
\end{figure}

Thus, there is strong motivation to study early SN emission in the UV.
The bolometric luminosity of the early UV emission from SNe remains nearly constant, while the temperature of the cooling, 
expanding gas declines with time. In any given band, the measured flux will rise as the peak of the emitted spectrum 
cools and approaches the band center, reaching maximum when the spectral peak is within the band, then declines 
as further cooling drives the emission peak to redder wavelengths (Fig.~\ref{SCbasics}; see an animated version on
\url{http://www.weizmann.ac.il/astrophysics/ultrasat/animations/ganot14a.gif}). 
The rate of cooling (and thus the time 
it takes for the flux to peak in a given band) depends on the stellar radius and the composition of the envelope 
which determines the opacity. For supergiant explosions with thick hydrogen envelopes, the opacity is known 
(Thomson scattering) and time independent, so the radius is straightforwardly inferred (Fig.~\ref{getR}).
For evolved (e.g., W-R) stars the opacity is a function of the surface composition (mass fraction of He, C and O).
RW11 show that, given a well-sampled UV light curve, one can infer the stellar radius and 
constrain the surface composition (Fig.~\ref{getZ}). 

Model calculations such as those of RW11 above assume standard massive star models 
and this leads to the prediction of a nearly-constant shock-cooling luminosity. However, some recent 
works hint that at least some stars undergo violent pre-explosion evolution, e.g., eruptive mass loss 
(e.g., Pastorello et a. 2007; Ofek et al. 2013, 2014a, Gal-Yam et al. 2014) and thus their pre-explosion density distribution may 
strongly deviate from standard models. In this case the shock-cooling luminosity will not be constant. 

However, RW11 also show that by combining UV and optical 
data one can determine the exact extinction towards an event and, correcting for it, measure both the 
temperature evolution and the radius without any assumptions regarding 
a constant shock-cooling 
luminosity. For such events, the luminosity and temperature evolution extracted from the UV+optical data will then 
measure the non-standard density profile, mapping recent pre-explosion mass loss and the 
physics of the final stages of stellar evolution. 

\begin{figure}
\includegraphics[width=17cm]{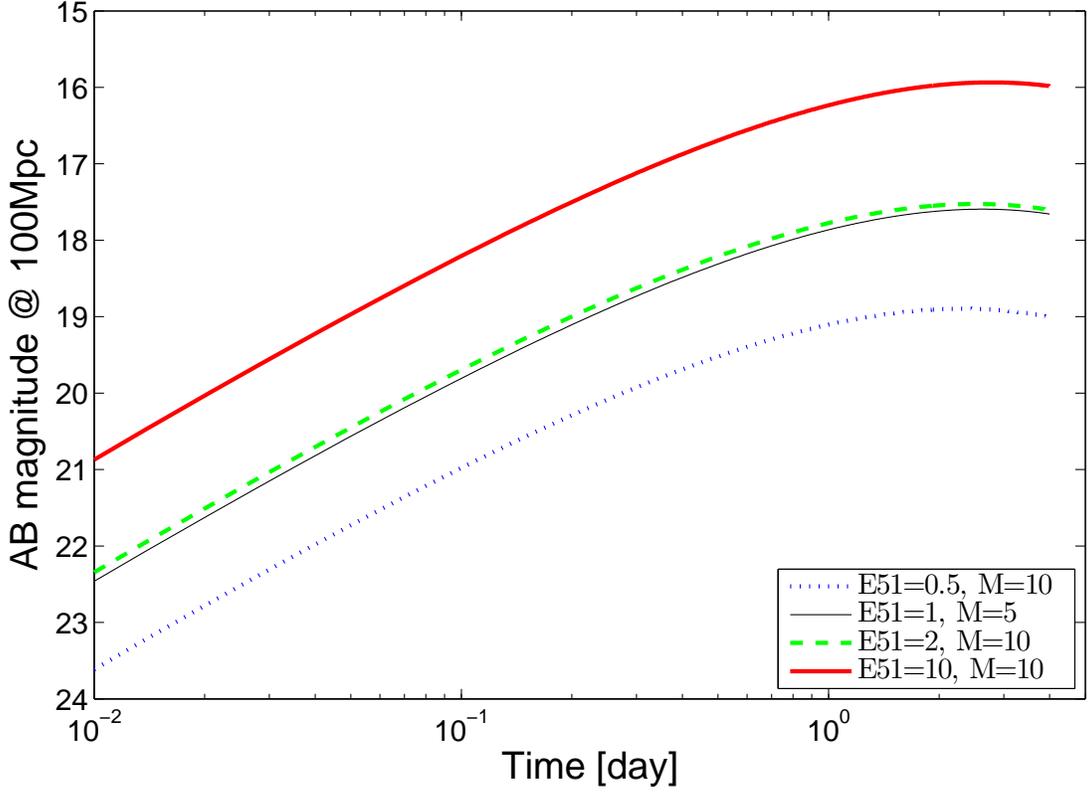}
\caption{Model RW11 SN NUV light curves for RSG stars with identical radii ($500$\,R$_{\odot}$) and several values of the
explosion energy E (in units of E51$=10^{51}$\,erg) and ejected mass M (in solar mass). 
The light curve shapes are identical (since these depend only on the radii and composition)
while the luminosity is a function of the ratio E/M (e.g., compare the dashed green and thin black curves). Assuming the extinction toward an event
has been measured via the combination of UV and optical observations (RW11), one can use the luminosity to measure
the value of E/M. Additional optical observations over longer time scales can constrain the ejected mass and allow to independently 
 infer both the explosion energy and the ejected mass separately (e.g., Barbarino et al. 2014).}
\label{getEM}
\end{figure}

In all cases, the extinction-corrected absolute luminosity evolution 
can be used to derive the energy per unit mass in the exploding ejecta (E/M), yet another vital
constraint on the explosion (Fig.~\ref{getEM}). The full route from UV light curves to physical stellar parameters has been 
demonstrated for (the only) three SN events with useful data (Types II and Ib; Schawinski et al. 2008; Gezari et al.
2008; Soderberg et al. 2008) by RW11. 

\begin{figure}
\includegraphics[width=15cm]{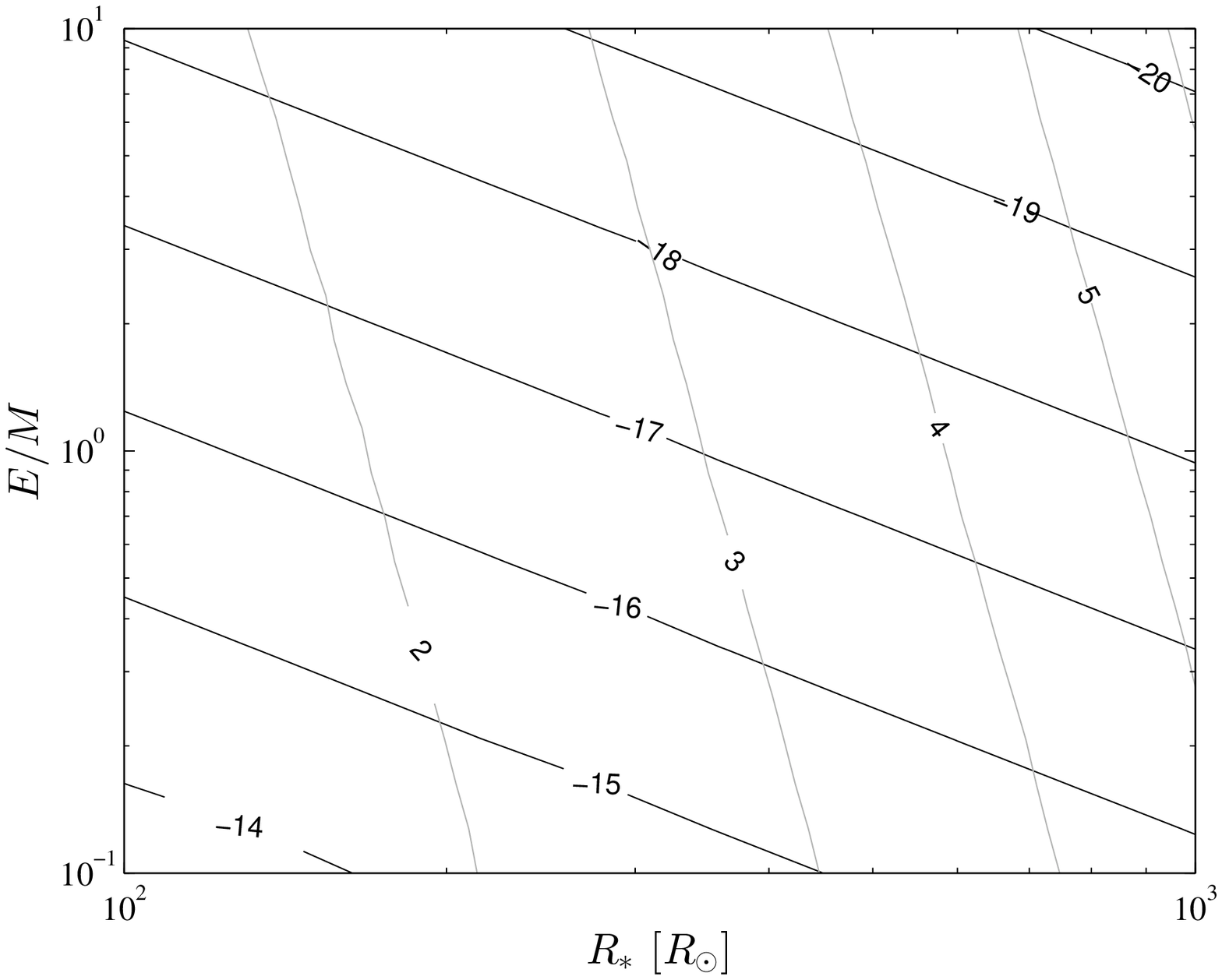}
\caption{Lines of constant peak absolute UV magnitude in the ULTRASAT band (black) and rise time in days (thin grey) predicted
by RW11 models for RSG explosions as a function of the stellar radius R and explosion mass E/M. As can be seen, the near-orthogonality of these lines allows to simply deduce the underlying parameters from the observed peak and rise time values.}
\label{RnEvsPeaknRT}
\end{figure}

Fig.~\ref{RnEvsPeaknRT} shows how the progenitor radius and the explosion energy
per unit mass can be directly derived from measurements of the UV peak absolute magnitude and rise time. Useful 
formulae to connect the observed parameters to RW11 model parameters are provided below, for radii R$_{*}$ measured
in units of the solar radius R$_{\odot}$, energy E in units of $10^{51}$\,erg and normalized to ejecta masses
of $10$\,M$_{\odot}$. We provide formulae for the absolute magnitude in the ULTRASAT band (M$_{peak}^{USAT}$),
as well as for the {\it Swift} UVW1 and UVW2 bands (M$_{peak}^{UVW1}$, M$_{peak}^{UVW2}$). The rise time t$_{rise}$ 
is defined as the time in days it takes the UV magnitude to rise by 1 magnitude to peak.   

\begin{equation}
M_{peak}^{USAT} = -11.237 -2.278 log_{10}(R_{*}) -2.276 log_{10}(E)
\end{equation}
\begin{equation}
log_{10}(t_{rise}) = -0.934 +0.555 log_{10}(R_{*}) +0.060 log_{10}(E)
\end{equation}

\begin{equation}
M_{peak}^{UVW1} = -11.285 -2.278 log_{10}(R_{*}) -2.276 log_{10}(E)
\end{equation}
\begin{equation}
log_{10}(t_{rise}) = -0.873 +0.557 log_{10}(R_{*}) +0.060 log_{10}(E)
\end{equation}

\begin{equation}
M_{peak}^{UVW2} = -11.283 -2.278 log_{10}(R_{*}) -2.276 log_{10}(E)
\end{equation}
\begin{equation}
log_{10}(t_{rise}) = -1.043 +0.554 log_{10}(R_{*}) +0.060 log_{10}(E)
\end{equation}

Early UV emission is thus a powerful way to study the progenitor properties of SNe, motivating efforts
to measure it systematically for a large sample of SN events by wide-field UV surveys. We will now provide estimates of the expected SN detection rates by such surveys using observations
and theory.

\section{The GALEX/PTF wide-field, shallow UV variability survey}

\subsection{Survey description}

We conducted a UV wide-field transient survey during a nine week period from 2012 May 24 through 2012 July 28.
This survey used the GALEX NUV camera to cover a total area of 600 square degrees. 
Operating in scanning mode, the GALEX NUV camera observed strips of sky in a drift-scan mode with
an effective  average integration time of $80$\,s, reaching a NUV limiting magnitude of $20.6$\,mag AB. 
Each strip was visited once every 3 days. In parallel, we observed the same area with the 
Palomar Transient Factory (PTF; Law et al. 2009; Rau et al. 2009) in $r$-band, reaching a
limiting magnitude of R$\sim21$\,mag AB with a cadence of 2 days, weather permitting. 

The main scientific goals of this survey were to study the early UV emission from SNe (this 
work; Ganot et al. 2015, in prepartion), AGN variability, stellar activity (flares) and white
dwarf variability. We estimate the completeness of this survey to SNe exploding in the FOV and 
above the limiting magnitude (at some time) at $50\%$ mainly due to the combined effects of GALEX
data loss due to failed downlink and image corruption (about $20\%$), 
PTF weather losses and PTF survey incompleteness ($60\%$ compounded). A full description of 
this survey, its completeness and its results will be presented in a series of forthcoming papers.

\subsection{Detected SN sample}

In this initial work we limit our analysis to the sample of spectroscopically-confirmed SNe detected with PTF 
within the GALEX field of view during the survey period. The SN sample includes 33
Type Ia SNe that will be presented elsewhere as well as 10 core-collapse SNe.
We list these core-collapse SNe in Table~\ref{SNtable}, and review their properties below. 
Interestingly, our survey also detected a distant superluminous SN of Type II (SLSN-II; Gal-Yam 2012)
at $z=0.275$. This remarkable event (PTF12gwu; Fig.~\ref{SLSNfig}) will be the subject of a separate publication.
All of these events were spectroscopically classified as part of PTF operations and redshifts were measured
from host galaxy lines (except for a single case, PTF12fkp, where the redshift is determined at lower
accuracy from the SN lines). 

\begin{figure}
\includegraphics[width=16cm]{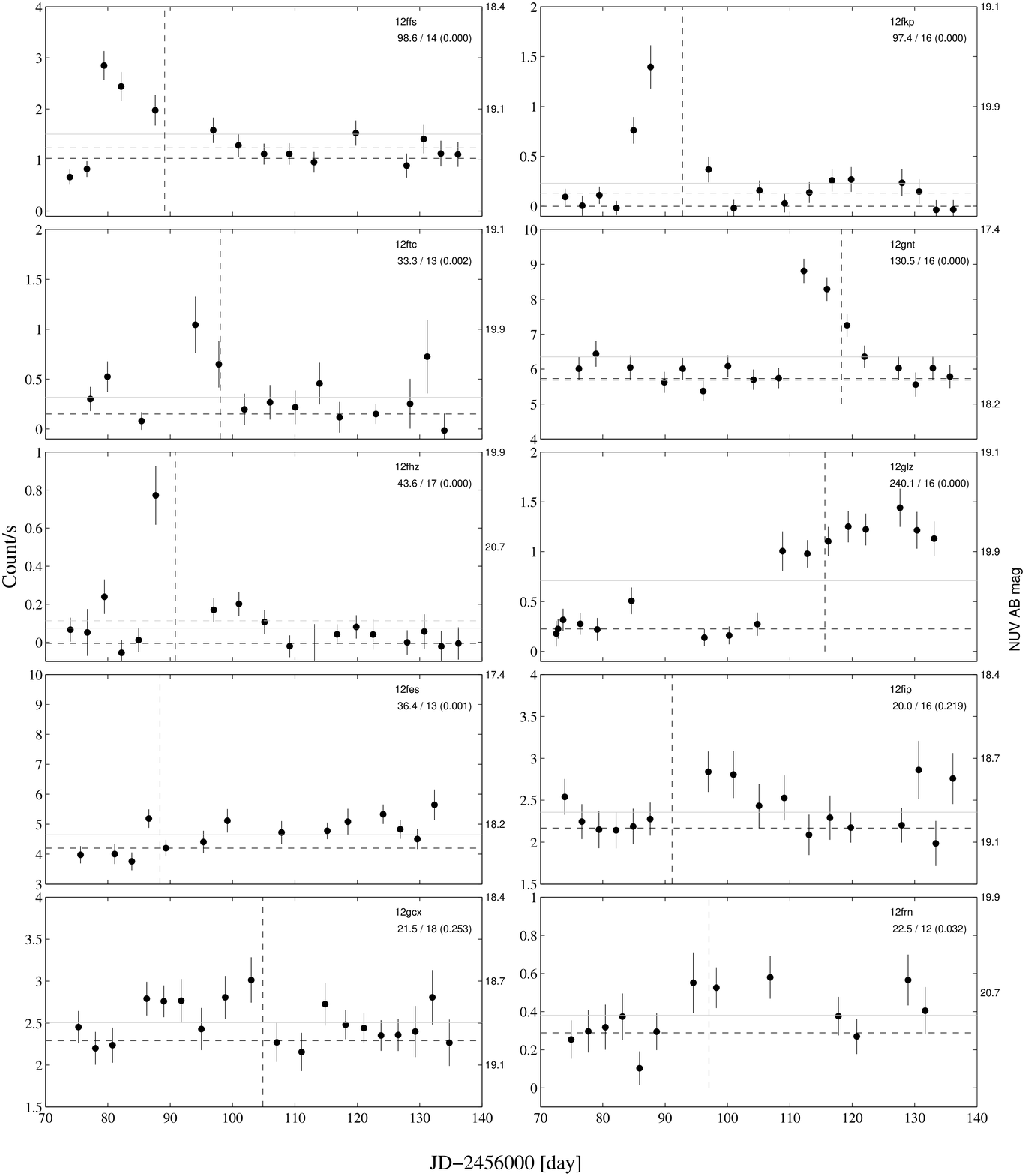}
\caption{GALEX light curves of the 10 PTF spectroscopically confirmed core-collapse SNe (solid circles). 
PTF discovery dates are marked with vertical lines. Blue dotted horizontal lines
indicate the flux level measured at these locations in pre-explosion GALEX data obtained prior to the 
start of our experiment; dashed horizontal lines are
the quiescent flux level indicated from our own GALEX data (see text). Below each object name 
we report the $\chi^2$ per degrees of freedom obtained when fitting the data with
a constant flux level, and in parenthesis the false alarm probability (FAP). 
Values below FAP$=0.01$ are marked as zero. Six events show clear
UV flares (top panels). Only four objects show no significant 
UV flare emission (bottom panels; see text).}
\label{LCfig}
\end{figure}

\begin{figure}
\includegraphics[width=17cm]{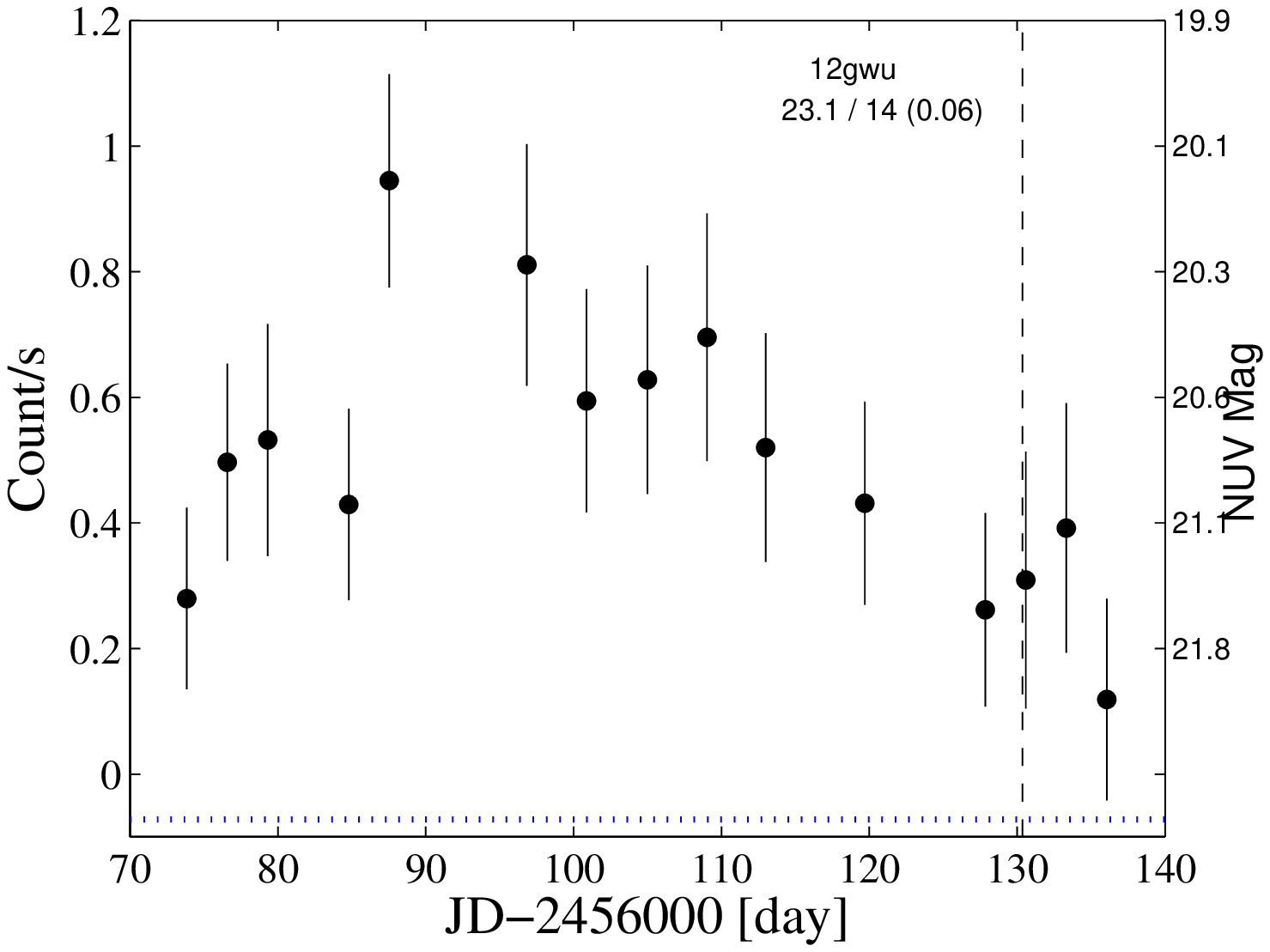}
\caption{The GALEX light curve of PTF12gwu, a SLSN-II. Symbols and curves are the same as in Fig.~\ref{LCfig}.
This event probably occurred around or shortly prior to the start of our GALEX/PTF experiment, which detected 
a very luminous and prolonged emission from this event. Analysis of these data will be presented in a forthcoming
publication.}
\label{SLSNfig}
\end{figure}

We show the GALEX NUV light curves of the ten core-collapse SNe in Fig.~\ref{LCfig}. GALEX UV photometry was
measured at the PTF SN locations using custom aperature photometry routines (Ofek 2014). We used an aperture of 5 pixels ($7.5\arcsec$). The sky was measured in an annulus with inner and outer radii of 20 and 50 pixels, and we used a zero point of $20.08$\,mag and an aperture correction of $0.18$\,mag for
the GALEX NUV camera (Morrissey et al. 2007). 
The photometry is marked by solid circles
with 1$\sigma$ error bars.  
PTF discovery dates are marked with vertical lines. Blue dotted horizontal lines
indicate the flux level measured at these locations in pre-explosion GALEX data obtained prior to the 
start of our experiment. When such past imaging is not in hand, we indicate with dashed horizontal lines
the quiescent flux level as measured from our GALEX data (the $25\%$ precentile flux level, 
to avoid contamination by the SN flare emission). To assess detection significance we calculated
the $\chi^2$ and number of degrees of freedom obtained when fitting the data with
a constant flux level, noted below each object name in Fig.~\ref{LCfig}, where we also 
report in parenthesis the resulting false positive probability (FPA). Six events show clear
UV flares (top panels; low FPA). Only four objects show no significant 
UV flare emission (bottom panels). Of those four events, two (PTF12fip and PTF12gcx) are consistent with 
a constant flux (solid grey line). Two other events (PTF12fes and PTF12frn) are inconsistent with
a constant flux (low FPA) but show no clear flare-like structure. We conclude that 6 GALEX events are
robustly detected. 

Of those six, PTF photometry and spectroscopy indicates that they all are Type II SNe (2 SNe II-P, 1 intermediate II-P/L, 1 II-L, 1 IIn and 1 IIb; Table~\ref{SNtable}). 
The mean redshift of the GALEX-detected sample, 
as well as of the entire set of core-collapse SNe is $z\sim0.07$. Interestingly, of the six GALEX-detected 
SNe only one occurred in a luminous host, while four are located in dwarf galaxies, only marginally detected in 
our GALEX NUV observations. This indicates that for NUV-detected core-collapse SNe, the host galaxy light
contribution to the background is typically negligible. Most events are detected only during a small number
of GALEX epochs (1-3) around their PTF discovery date, while the single detected Type IIn SN shows a 
prolonged period of UV luminosity extending beyond the duration of our survey period. Interestingly,
in all cases the first UV detections occur prior to the optical discovery by PTF, elucidating the superiority
of the UV over the optical for early SN studies. 

We note that one out of these 6 events (PTF12glz) is a luminous Type IIn
SN and displays a prolonged UV emission.
The light curves of such events were suggested
to be powered, at early times, via the explosion shock 
breaking out from a spatially extended opaque wind, rather than from the surface of a star
(Ofek et al. 2010) and the decaying part is presumably due to the conversion
of kinetic energy to optical luminosity (see also Chevalier \& Irwin 2011; 2012;
Balberg \& Loeb 2011; Moriya \& Tominaga 2012; Ginzburg \& Balberg 2012; Ofek et al. 2014b 
and Svirski \& Nakar 2014).
Such events
are relatively rare in the volumetric sense, but their detectability to larger distances compensates for this
in flux-limited surveys. Out results indicate that in shallow UV surveys such events will constitute $15-20\%$
of the sample.    
An extreme such case are SLSNe that are so UV-luminous that they are detected over a huge volume, and
may have similar detection rates (by number). 

A brief report
about these events was presented in Barlow et al. (2013), and a detailed analysis will be presented
in Ganot et al. (2015, in preparation) and additional future publications.

\begin{table}[h]
\caption{Sample of core-collapse SNe detected by the GALEX/PTF experiment}
\begin{threeparttable}
\centering
\begin{tabular}{l l l l l l}
\hline
PTF name & RA & Dec & Redshift & Type & PTF discovery date \\
\hline
PTF12ffs & 14:42:07.33 & +09:20:29.8 & 0.0511 & SN II\tnote{a} & June 10, 2012 \\
PTF12fhz & 15:18:20.09 & +10:56:42.7 & 0.0987 & SN IIb & June 12, 2012 \\
PTF12fkp & 14:46:54.81 & +10:31:26.4 & 0.12 & SN II-L & June 14, 2012 \\
PTF12ftc & 15:05:01.88 & +20:05:54.6 & 0.0732 & SN II-P & June 19, 2012 \\
PTF12glz & 15:54:53.04 & +03:32:07.5 & 0.0799 & SN IIn & July 7, 2012 \\
PTF12gnt& 17:27:47.30 & +26:51:22.1 & 0.029 & SN II-P & July 9, 2012 \\
\hline
PTF12fes & 16:00:35.13 & +15:41:03.5 & 0.0359 & SN Ib & June 9, 2012 \\
PTF12fip & 15:00:51.04 & +09:20:25.1 & 0.034 & SN II-P & June 12, 2012 \\
PTF12frn & 16:22:00.16 & +32:09:38.9 & 0.136 & SN IIn & June 18, 2012 \\
PTF12gcx & 15:44:17.32 & +09:57:43.1 & 0.045 & SN II\tnote{b} & June 26, 2012 \\
\hline
\end{tabular}
\begin{tablenotes}
            \item[a] A bright SN II with a light curve intermediate between SNe II-P and II-L
            \item[b] A bright SN II with a very long rise time, similar to SN 1998A (Pastorello et al. 2005), SN 2000cb 
(Kleiser et al 2011) and SNe 2005ci and 2005dp (Arcavi et al. 2012).
\end{tablenotes}
\end{threeparttable}
\label{SNtable}
\end{table}

\section{Theoretical estimates of early UV emission from SNe}

\subsection{Light curve models and comparison with data}

We have calculate theoretical early UV light curves for SNe in the following manner.
We use the analytic formalism of Rabinak \& Waxman (2011; RW11), that has been tested
against numerical simulations and self-similar solutions and describes available
observations well ($\S~2$). Other analytical models (NS10; Chevalier 1992) are broadly similar
and using those instead does not alter our derived detection rates.  
We are careful to correct the typographical error appearing in the RW11 formulae 
according to the published Erratum (Rabinak \& Waxman 2013).

Our calculations include the following steps. First, we calculate the RW11 bolometric luminosity for a set
of progenitor parameters (radius R$_{*}$; explosion energy E and ejected mass M). The model parameter $f_{\rho}$
is set to its suggested value of $f_{\rho}=0.1$.
We use the Thomson opacity for
supergiant stars, and the prescriptions of RW11 for mixed He/C/O envelopes of W-R stars. 
We corrected the temperature up by a factor of 1.2 as suggested by RW11, to account for the fact that the color temperature
is set at Thomson optical depth above unity (see NS10 for an analytic approximation). Using 
the evolving radius and temperature, we then calculate black-body spectral curves
as a function of time. Convolving these spectra with a sensitivity curve for a given 
observational band (e.g., NUV or optical bands), we calculate the light curve in these bands via synthetic 
photometry (Ofek 2014). 

\begin{figure}
\includegraphics[width=17cm]{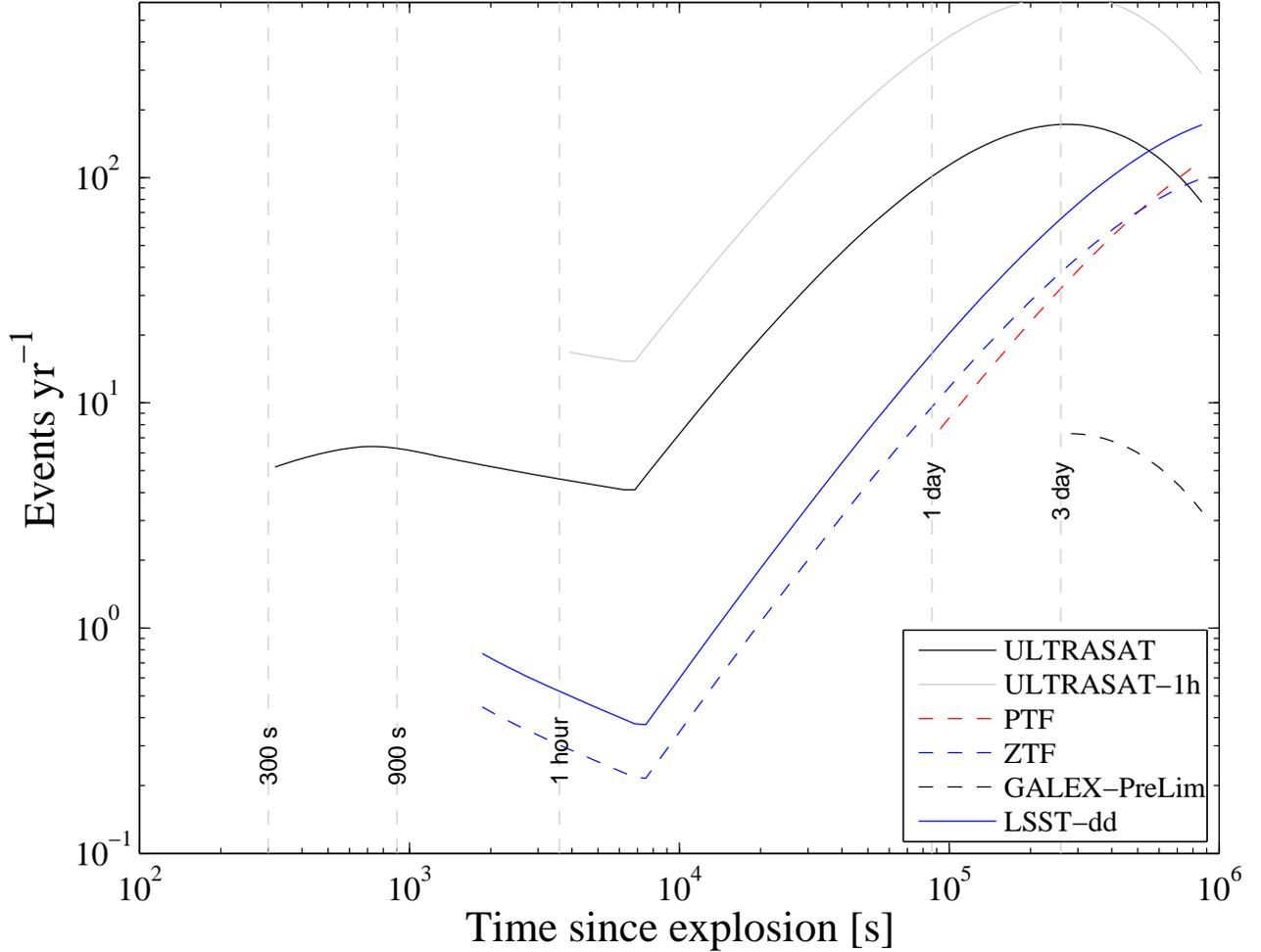}
\caption{Expected number of early UV SN detections for various surveys. We use the models of Sapir et al. (2013) at
early times, and RW with our fiducial parameters (R$_{*}=500$\,R$_{\odot}$, E$=10^{51}$\,erg and M$=10$\,M$_{\odot}$)
at later times. The resulting expected detection rates for several surveys (Table~\ref{predictiontable})
are plotted as a function of the time since explosion, starting with that expected for the survey
minimal cadence. The curves are not cumulative, i.e., they indicate how many events will be detected {\it at a given 
age} (and not {\it below} that age). The survey parameters are given in Table~\ref{predictiontable}).}
\label{detnums}
\end{figure}

To determine object detectability we now assume a distance
as well as a value for Galactic extinction, and calculate the flux from an object as a function of time since explosion, and the 
distance. This is then translated to a number of detections for a survey with a given sensitivity (depth) in a given band, 
and a given field of view (e.g., Fig.~\ref{detnums}), if we know the volumetric rate of the event in question.

\subsection{Volumetric rates}

We use the volumetric SN rates from Li et al. (2011). Using a sample of nearby SNe these authors have measured
rates for Type II SNe and Type Ib/c SNe of approximately $0.5\times10^{-4}$\,SN\,Mpc$^{-3}$\,yr$^{-1}$ and
$0.25\times10^{-4}$\,SN\,Mpc$^{-3}$\,yr$^{-1}$, respectively. Since Type II SNe typically result from 
red supergiant stars (Smartt 2009), we normalize our predictions for RSG explosions using the Type II SN rate.

In contrast, the nature of the progenitors of SNe Ib/c is still unclear. While some indications exist that these
arise from compact W-R stars (e.g., Corsi et al. 2012, Cao et al. 2013), other events appear to require
more extended progenitors with radii well above $10^{11}$\,cm (e.g., SN 2008D, RW11; SN 2006aj, Nakar \& Piro 2014).
We have thus aribitrarily assigned $50\%$ of the volumetric rate of SNe Ib/c to compact W-R progenitors, and
$50\%$ to more extended BSG-like stars. For optical/UV surveys the expected number of early detections 
of such compact stars are, in any case, small compared to the dominant population of red supergiant 
explosions.

\section{Results}

Combining our UV light curve models and the measured SN volumetric rates, we can predict the expected number of detections
of various progenitor explosions (RSG, BSG and W-R) as a function of progenitor parameters R$_{*}$, E and M. We begin
by comparing our predictions with the GALEX/PTF survey we have conducted ($\S~5.1$), and provide predictions
for other space ($\S~5.2$) and ground-based ($\S~5.3$) surveys. We discuss the fractions of SN types in wide-field
surveys in $\S~5.4$ based on observed PTF data. 
  
\subsection{Predicted rates for the GALEX/PTF experiment and fiducial progenitor parameters}

Using the procedure described above, we predict the expected number of early UV detections of Type II SNe
in our GALEX/PTF experiment assuming all of these result from RSG progenitors with a single set of fiducial
parameters. We set these to be R$_{*}=500$\,R$_{\odot}$, E$=10^{51}$\,erg and M$=10$\,M$_{\odot}$,
which agree with typical values for RSG radii and energy and mass estimates for Type II SNe. Using this set of 
fiducial parameters and the RW11 models, we predict that our survey should have detected 7 SNe (Table~\ref{predictiontable};
assuming it was $50\%$
complete, see above). Comparing this with the actual number of 6 detections ($\S~3$; Fig.~\ref{LCfig}), we find good
agreement with the predictions given the small numbers involved. We conclude that 
using the set of fiducial RSG model parameters and RW11 models to predict early UV SN detection numbers is validated by 
our GALEX/PTF experiment.  

We calculate the number of BSG and W-R explosions using the parameters 
R$_{* BSG}=50$\,R$_{\odot}$, E$_{BSG}=2\times10^{51}$\,erg and M$_{BSG}=10$\,M$_{\odot}$ and for W-R
stars R$_{* WR}=10$\,R$_{\odot}$, E$_{WR}=2\times10^{51}$\,erg and M$_{WR}=10$\,M$_{\odot}$ and the
rates from $\S~4.2$. Even
with these rather optimistic parameters (high E and large R), 
the predicted number of BSG and W-R explosion detections within the GALEX/PTF experiment is small ($<1$). This is consistent with
our non detection of early UV emission from SNe Ib/c (or the peculiar SN II PTF12gcx, which we suspect may have had a 
BSG progenitor). We retain these as fiducial parameters for predictions, but note these are not constrained by our
observations.
 
\subsection{Predictions for ULTRASAT}

Having in hand a set of calibrated fiducial progenitor parameters for RSGs 
(R$_{*}=500$\,R$_{\odot}$, E$=10^{51}$\,erg, M$=10$\,M$_{\odot}$; RW11 models) 
that have been validated by reproducing our GALEX observations,
we can now predict the expected rates for the proposed ULTRASAT mission. For BSG and W-R explosions
our parameters are poorly constrained by data, so any predictions are tentative, but the rate is expected to be dominated
by RSG explosions (and it is). This wide-field UV space explorer
has been described in Sagiv et al. (2014), and here we use its current technical formulation, a field of view of 210 square degrees and 
a 5$\sigma$ limiting sensitivity of $21.5$\,mag AB in $900$\,s integration in the NUV ($220-280$\,nm band). 

As can be seen in Table~\ref{predictiontable},
ULTRASAT is predicted to discover the early shock-cooling emission from no less than 110 events per year. Of these, the large majority
(100) are expected to be due to RSG explosions. A handful of events (formally 6 per year) are expected to be detected during the shock-breakout 
phase ($<1$\,hour after explosion) but we consider this number only as a tentative estimate since the theory of SN emission at this
phase has not been tested observationally yet. We note that this prediction does not account for extinction of these SNe in their hosts, but such extinction will not affect the rate prediction. The reason is that we chose our fiducial RSG parameters to match
the observed GALEX/PTF rate. If we include an arbitrary mean extinction in our modelling (reducing the expected number
in the GALEX experiment), this would drive the RSG parameters towards values with brighter UV flares (larger R or higher E/M)
to exactly compensate and return the expected rate to its observed value. 
The effects of extinction thus cancel out and our predicted rates remain the same. 

\begin{figure}
\includegraphics[width=17cm]{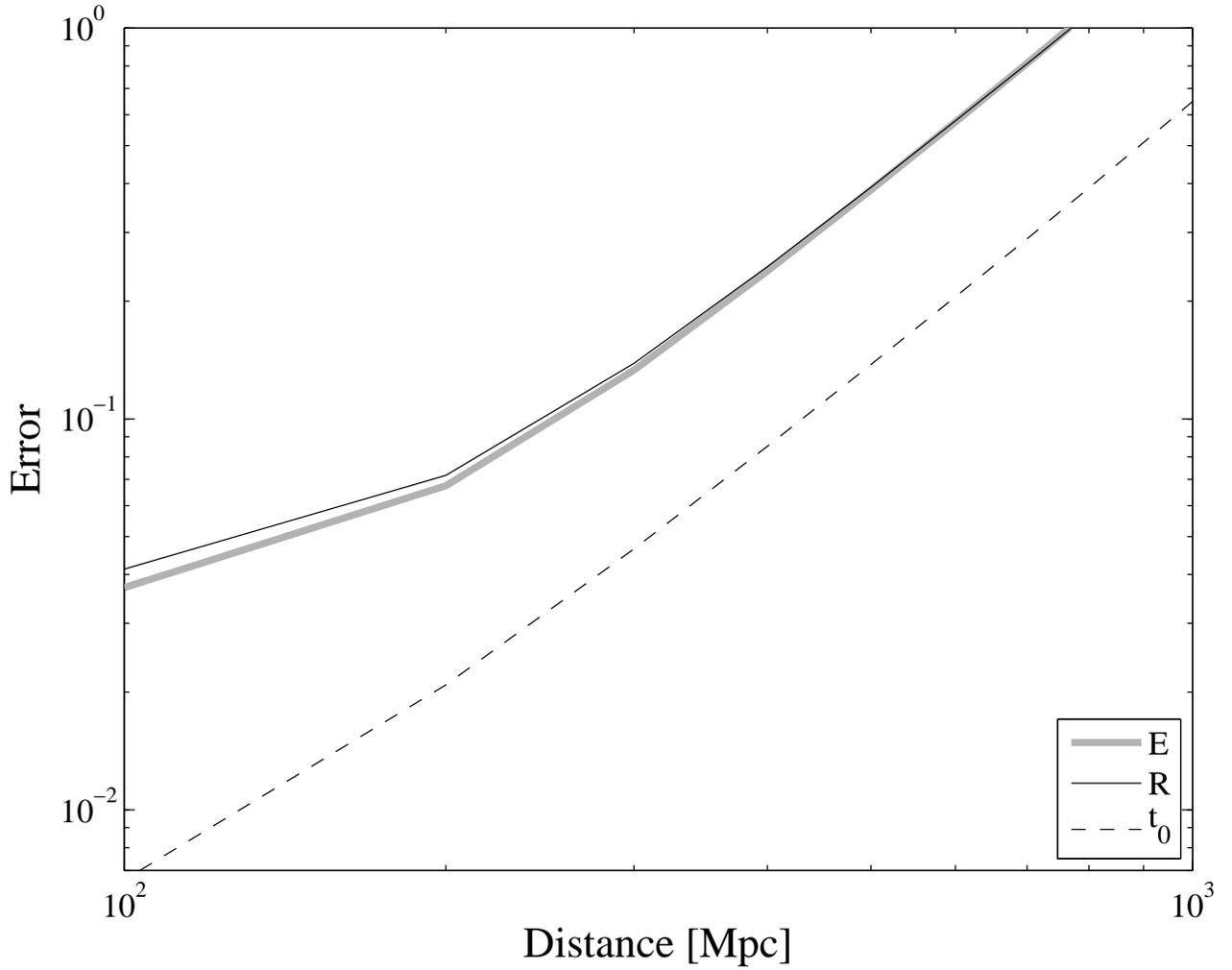}
\caption{The expected fractional errors on supernova progenitor and explosion parameters derived from covariance
matrix analysis for RW11 models with our fiducial RSG progenitor parameters, see text for details.}
\label{errorfig}
\end{figure}

We can estimate the expected accuracy with which we can derive progenitor and explosion parameters from ULTRASAT
data in the following manner. For out fiducial RW11 models, we calculate the covariance matrix taking $R_*$, $E/M$ and the
explosion time $t_0$ as free parameters, Poisson errors appropriate for the distance of a given event, its expected luminosity
and the ULTRASAT sensitivity (limiting magnitude of M$_{NUV}=21.5$\,AB Mag during $900$\,s integrations), as well as $3\%$
systematic errors. The square root of the diagonal elements of the covariance matrix are reported in Fig.~\ref{errorfig}. 
As can be seen, ULTRASAT will provide accurate measurements of these parameters (to below $10\%$ out to $200$\,Mpc).

Another simple, robust and extinction-free lower limit on the ULTRASAT detection rate is obtained by scaling our GALEX results. Our GALEX experiment 
detected N$_{GALEX}=6$\,events in t=2\,months, using a survey area of $\Omega=600$\,deg$^{2}$, had a 5$\sigma$ flux sensitivity of S$_{NUV}=20.6$\,mag AB, and covered a sky area with $\delta$A$_{NUV}=0.31$\,mag higher NUV extinction compared to the ULTRASAT NUV-optimized fields. Thus,
direct scaling to the ULTRASAT yearly yield would give N$_{ULTRASAT}=$N$_{GALEX} \times (1$\,year / 2\,months) $\times (210/600) \times$ (S$_{ULTRASAT}-$S$_{GALEX}+\delta$A$_{NUV}$)$^{3/2}$ which yields N$_{ULTRASAT}=71$\,events\,y$^{-1}$. This lower limit is based purely
on NUV-detected SNe so it accounts for all sources of extinction. It is also a conservative lower limit since it uses the rate from the GALEX/PTF survey
which had a UV cadence of 3\,d, compared to the expected ULTRASAT cadence of 900\,s. Many short events likely missed by our GALEX/PTF
search could be detected by ULTRASAT. Finally, this assumes a $100\%$ efficiency for our GALEX/PTF experiment, while in reality its completeness
was no more than $\sim50\%$ (see above). Correcting just for this factor, our lower limit is in good agreement with the theoretical prediction. The formal error on this lower limit is due to Poisson small number
statistics related to the GALEX detection number (6). At $95\%$ confidence, this error is by a factor
of 2 or less (Gehrels 1986).

We conclude that ULTRASAT is absolutely guaranteed to find $>70$ explosions of large RSG stars per year using this direct scaling from our GALEX observations, and that this number is with high confidence twice as large. The predicted detection rate $R$ for other
UV missions operating in a similar wavelength range can be estimated using simple scaling according to field of view $\Omega$
and limiting flux $S$, $R \propto \Omega \times S^{3/2}$. 

An interesting final point is that our GALEX/PTF experiment discovered one superluminous SN (SLSN-II) in two months. 
Using the same scaling above for ULTRASAT indicates this mission will likely detect $~\sim10$ SLSNe per year. 
These will be quite unique in having been discovered early and 
having UV coverage, which is crucial in order to shed light on the progenitors and power sources of these most energetic 
and UV-bright objects.

\subsection{Predictions for other surveys}

The same models we apply above can also be used to predict the expected early SN discovery rate for other experiments, in particular
ground-based optical surveys. We note that these surveys would not be able to carry out the science investigation motivated in $\S~2$
since, as explained above, it requires early UV data. However, these ground-based surveys could conceivably trigger UV follow-up observations from space
(e.g., by {\it Swift} or even {\it HST}) that will allow progenitor and SN physics to be extracted from early data, at least for a few
objects.
 
The iPTF survey at Palomar Observatory is operating the PTF survey camera and has demonstrated its ability to quickly discover SNe and trigger
space-based UV follow-up (Gal-Yam et al. 2011). We calculate the expected number of events for an iPTF survey covering $1000$\,deg$^{2}$
with a nightly cadence in $r$-band (Table~\ref{predictiontable}). iPTF has a lunation-averaged depth of $r=20.6$\,mag, and we have assumed
a $25\%$ temporal efficiency (including loss due to daytime and weather). The predicted yearly yield (9 events) is consistent with 
iPTF detections of several SNe at ages $<1$\,d so far (e.g., Gal-Yam et al. 2014).    

Next, we consider the coming Zwicky Transient Facility (ZTF) that will be using an even larger camera mounted on the same telescope
at Palomar Observatory. While the observing strategy of ZTF has not been finalized yet, we consider here a $g$-band survey of $2100$\,deg$^{2}$
with a cadence of $0.5$\,h. We assume this survey will use $50\%$ of the ZTF time, and the same temporal efficiency as above.  With
a dozen or so events per year, all securely detected at $<1$\,day, ZTF, hopefully coupled with {\it Swift}, will be able to provide the first
few examples of the science expected from ULTRASAT.
  
Finally, it is interesting to consider what the Large Synoptic Survey Telescope (LSST) could achieve. Assuming that, as part of a ``deep drilling''
experiment, LSST will observe a single field of view at any given time repeatedly every $0.5$\,h in the $g$-band with a lunation-averaged depth
of $g=24.2$\,mag and a temporal efficiency as above, LSST will be able to perhaps detect less than 1 SN per year within 1h of explosion, and
about 20 events within 1 day, still well below the expected performance of ULTRASAT. In addition, these events will be typically
distant and hence faint, and difficult to follow-up (e.g., spectroscopically).  

\begin{table}[h]
\caption{Predicted SN explosion detection numbers by various surveys}
\begin{threeparttable}
\centering
\begin{tabular}{l l l l l l l l l l}
\hline
Survey & Band & Cadence & FOV [deg$^2$] &  \multicolumn{6}{c}{Expected Number [SN yr$^{-1}$]}  \\
            &          &                &                      &\multicolumn{2}{c}{RSG}    & \multicolumn{2}{c}{BSG}  & \multicolumn{2}{c}{W-R}\\                
            &          &                &                      & $<1$\,h  & $<1$\,d & $<1$\,h  & $<1$\,d & $<1$\,h  & $<1$\,d \\
\hline
GALEX/PTF          & NUV & 3\,d           & 600 & 0  & 42\tnote{a} & 0 & 0 & 0 & 0\\
\hline
ULTRASAT           & NUV & $900$\,s    & 210 & 6 & 100 & 1 & 8 & 0 & 4\\
ULTRASAT           & NUV & $3600$\,s    & 210 & 20 & 380 & 3 & 30 & 1 & 13\\
\hline
iPTF \tnote{b}     & $r$  & 1\,d            & 1000 & 0 & 6 & 0 & 2 & 0 & 1\\
ZTF\tnote{c}      & $g$  & 0.5\,h        & 2100 & 0 & 9 & 0 & 2 & 0 & 1\\
LSST\tnote{d}     & $g$  & 0.5\,h          & 9.6    & 0 & 17 & 0 & 3 & 0 & 2\\
\hline
\end{tabular}
\begin{tablenotes}
\item[a] For our GALEX/PTF Pre-LIM experiement, we report here the expected number within 3\,d (not 1\,d) to match its low actual cadence. As the survey ran for 2\,m (1/6\,yr), the expected number of SNe from RSG explosions for the actual experiment is 42/6=7\,events.
\item[b] Assumed temporal efficiency of $25\%$ (including loss due to daytime and average weather) and lunation-averaged depth of 20.6\,mag.
\item[c] $25\%$ temporal efficiency as above, average depth 20.4\,mag, and $50\%$ survey time spent in g-band. 
\item[d] Assumed the following for the LSST deep-drilling project: 1 LSST field observed at any given time,  $25\%$ 
temporal efficieny as above, g=24.2\,mag lunation-averaged depth.
\end{tablenotes}
\end{threeparttable}
\label{predictiontable}
\end{table}

\subsection{Estimated SN fractions in wide-field surveys}

In view of the complexity of massive stars and the resulting diversity of their explosive
core-collapse SN outcomes, an important aspect in the design of a survey to systematically 
study early emission from massive star explosions is the number of different SN types one expects
to obtain significant data about. To assess this for our own survey, as well as future programs
such as ULTRASAT (Sagiv et al. 2014), we use the large sample of spectroscopically-confirmed
SNe from the PTF survey (Law et al. 2009; Rau et al. 2009; Arcavi et al. 2010). The final sample
from PTF (2009-2012) includes 484 events. This sample is suitable for our study as it comes from
a relatively shallow survey with a depth identical to the GALEX/PTF survey (by definition) and similar
to that predicted for ULTRASAT. The survey is also untargeted (it is not focussed on known catalogued 
galaxies that are typically biased towards more massive and metal-rich objects).

The fractions of SNe of different types from the PTF flux-limited sample are reported in Table~\ref{fractiontable}.
The separation of the common class of Type II SNe into photometric subclasses (II-P and II-L) should not
be regarded as final, and is, in any case, controversial (e.g., Arcavi et al. 2012; Anderson et al. 2013; Faran et al. 2014a,b).
The size of the PTF sample allows estimates of the observed fractions of even rare classes (e.g., Ic-BL and II-pec) with
reasonable accuracy. 

Table~\ref{fractiontable} also provide estimates for a fiducial sample of 100 events,
as well as the $95\%$ confidence lower-limit on the expected number of SNe from each class
in this fiducial sample. We find that all SN types except for the rare Type II-pec events are
expected to be detected in samples of size 100 events or larger. 

Next, we estimate the expected yield of our GALEX/PTF experiment and compare it with our actual 
findings (right columns of Table~\ref{fractiontable}). We find excellent agreement even for the small
numbers in questions. We counted PTF12ffs as a Type II-P event, but including it instead in the
II-L class would not significantly alter this result in view of the small SN numbers.  

\begin{table}[h]
\caption{SN fractions from 484 PTF core-collapse SNe and predictions from other surveys}
\begin{threeparttable}
\centering
\begin{tabular}{l l l l l ll}
\hline
SN Type & PTF number & Fraction & Expected \#  & Minimum \#                 & Expected          & Actual \\
              &                     & of total  & (per 100)       & ($95\%$ c.l.)\tnote{a}  & (per 10)         & GALEX/PTF \\
\hline
II-P        & 193              & 40\%     & 40                  & 30                                  & 4                     & 4  \\
II-L        & 70                & 14\%     & 14                  & 8                                    & 1                     & 1 \\
IIb         & 24                & 5\%       & 5                    & 2                                    & 1                     & 1 \\
IIn         & 91                & 19\%     & 19                  & 12                                  & 2                     & 2 \\
II-pec\tnote{b}    & 5                  & 1\%       & 1                    & 0                                    & 0                     & 1\\
Ib          & 34                & 7\%       & 7                    & 3                                     & 1                     & 1 \\
Ic           & 49                & 10\%     & 10                  & 5                                    & 1                     & 0 \\
Ic-BL      & 18                & 4\%       & 4                    & 1                                    & 0                     & 0 \\
\hline
\end{tabular}
\begin{tablenotes}
\item[a] Minimum number expected per 100 events at a confidence level of $95\%$ calculated using
small-number Poisson statistics (Gehrels 1986).
\item[b] Peculiar Type II SNe with very long rise times, similar to SN 1987A or PTF12gcx (see note to Table~\ref{SNtable})
\end{tablenotes}
\end{threeparttable}
\label{fractiontable}
\end{table}

\section{Conclusions}

Motivated by the scientific promise of early UV observations of SNe, we have conducted a GALEX/PTF survey for such events
that detected six Type II SNe at ages $<3$\,d. We develop a theoretical framework to predict the number of early UV SN detections
in general surveys, using theoretical UV light curves that fit existing data well, combined with measured volumetric SN rates. 
We find that adopting a set of reasonable physical parameters for exploding RSG SN progenitors 
(R$_{*}=500$\,R$_{\odot}$, E$=10^{51}$\,erg and M$=10$\,M$_{\odot}$) fits our PTF/GALEX results well. We adopt these
parameters and predict the expected early UV SN detection numbers from the proposed ULTRASAT space mission, as well as several
ground-based surveys (Table~\ref{predictiontable}). We find that ULTRASAT is expected to discover $>100$ SNe per year in the 
UV, within 1\,day of explosion. A robust lower-limit
directly derived from the GALEX UV detection rates supports this estimate. Using SN Type statistics from PTF we show that such
a sample is likely to include examples of all common SN Types (Table~\ref{fractiontable}). We conclude that a space mission like ULTRASAT will be able
to comprehensively map the progenitor propoerties of SNe of all types (including radii and surface composition) and constrain
SN explosion physics, providing a compelling answer to the question of massive stellar death.    

\section*{Acknowledgments}

This research was supported by grants from the Israeli Space Agnecy (ISA) and the Ministry of Science, Technology and Space
(MOS).

\end{document}